\documentclass[9pt]{extarticle}

\usepackage{graphicx}
\usepackage{xspace}
\usepackage{authblk}
\usepackage[hidelinks]{hyperref}
\usepackage[left=1.5cm,right=1.5cm, top=2cm, bottom=2cm]{geometry}
\usepackage{multicol}
\usepackage{float}
\usepackage[tiny]{titlesec}
\usepackage{breakcites}

\titlespacing{\section}{0cm}{0.3cm}{0.1cm}

\graphicspath{ {./images/} }

\newcommand{\pcxg}{Portable-CELLxGENE\xspace}
\newcommand{\thetitle}{\Huge \pcxg: Standalone executables of CELLxGENE for easy installation}
\newcommand{\myemail}{george.hall@ucl.ac.uk}
\newcommand{\myname}{George T. Hall}
\newcommand{\ghurl}{\url{github.com/george-hall-ucl/portable-cellxgene}}

\begin{document}

\title{\thetitle}

\author[]{\myname}

\affil[]{\small Genetics and Genomic Medicine Department, \\ UCL Great Ormond Street Institute of Child Health, University College London, \\ 20 Guilford Street, WC1N 1DZ, London, United Kingdom \\ \vspace{0.2cm} \myemail}

\date{}

\noindent\par
\noindent\makebox[\textwidth][c]{%
    \begin{minipage}{15cm}
        \vspace{1.2cm}
        \maketitle
    \end{minipage}
    }

\vspace{0.5cm}

\begin{center}
\begin{minipage}[t]{14cm}
    Biologists who want to analyse their single-cell transcriptomics dataset
    must install and use specialist software via the command line. This is
    often impractical for non-bioinformaticians.  Whilst the popular CELLxGENE
    software provides an intuitive graphical interface to facilitate analysis
    outside the command line, its server-side installation and execution remain
    complex. A version that is easier to install and run would allow
    non-bioinformaticians to take advantage of this valuable tool without
    needing to use the command line. \pcxg is a standalone distribution of
    CELLxGENE that can be installed via a graphical interface.  It contains an
    easy-to-use extension of the CELLxGENE-Gateway Python package to allow the
    analysis of multiple datasets. \\

    \textbf{Availability and implementation:} Versions of \pcxg for Windows and
    MacOS, along with its source code, are available at \ghurl. It is licensed
    under the GNU General Public License v3.
\end{minipage}
\end{center}
\vspace{0.5cm}

\section{Statement of Need}

Software to analyse single-cell transcriptomics experiments requires time and expertise
to install and operate. Non-bioinformaticians must therefore often work in
close collaboration with bioinformaticians to carry out tasks such as labelling
cell types according to marker genes and assessing differential gene
expression. Research would be accelerated if non-bioinformaticians could
tackle these tasks alone. However, this is currently challenging due to the
lack of easy-to-install and easy-to-use software.

The CELLxGENE tool~\cite{cellxgene} is easy-to-use since it provides a
graphical user interface (GUI) through which single-cell datasets can be
analysed. The CELLxGENE-Gateway extension~\cite{cellxgene_gateway} allows it to
be used with multiple datasets, with the ability to create multiple versions of
cell-level annotations and gene-level gene sets. However, both tools remain
prohibitively complex for the non-bioinformatician to install since they
require both Python and the correct versions of many Python packages and other
software to be installed.  Whilst CELLxGENE instances can be hosted online --
thereby bypassing installation for the end user -- setting up the server
requires time, money, and specialist knowledge, in addition to the requirement
to remain online whilst conducting analysis.

Some other tools also provide a GUI for the analysis of single-cell
transcriptomics experiments, but each has shortcomings. The Galaxy
tool~\cite{galaxy} provides a GUI and allows for more complex analyses than
CELLxGENE, but this complexity may make it again unsuitable for
non-specialists. Whilst the Loupe browser developed by 10x Genomics
is easy to install and use, it can only operate on datasets saved in the
proprietary \texttt{loupe} format. These datasets are generally those that have
been processed by the 10x Genomics CellRanger software~\cite{cellranger} --
although datasets in other formats can now be converted to this
format~\cite{loupeR}. Nevertheless, the Loupe browser is not open source,
potentially limiting its future availability, extendibility, and utility in
some situations. \pcxg, on the other hand, is build on open-source software
using the well-maintained and widely-used \texttt{h5ad} file format.

\pcxg addresses the lack of an easy-to-install and easy-to-use tool by
providing standalone executable versions of the intuitive and powerful
CELLxGENE software. It will accelerate research by enabling
non-bioinformaticians to carry out simple analysis tasks independently from
bioinformaticians.

\clearpage

\section{Implementation}

\subsection{\pcxg}

\pcxg comprises a conda~\cite{conda} environment (containing Python, all
necessary Python packages, and all other software) along with a script to run
the tool. The conda environment incorporates CELLxGENE along with
CELLxGENE-Gateway, an extension which allows for the analysis and annotation of
multiple datasets.  Upon launch, a shell is started which activates the conda
environment. A window appears for the user to select the location of their
dataset. The launch script then sets environment variables to point to the
dataset and configure other CELLxGENE and CELLxGENE-Gateway options. Finally, a
CELLxGENE-Gateway session is started, and, after a small delay to allow for all
necessary processes to start, a browser window is opened to display the web
app.

\begin{figure}[htbp]
    \centering
    {
        \setlength{\fboxsep}{0pt}
        \setlength{\fboxrule}{1pt}
        \fbox{\includegraphics[width=\textwidth]{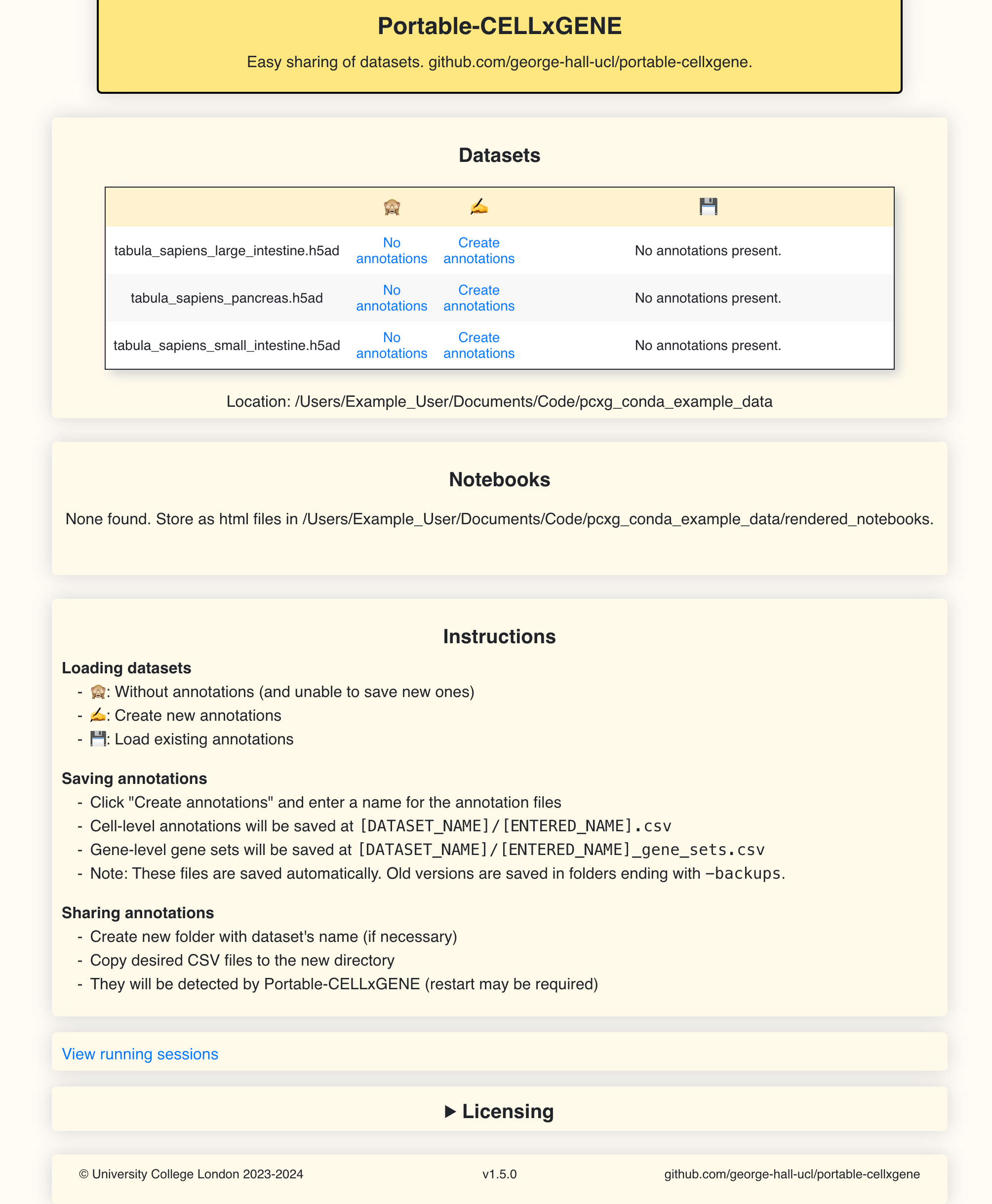}}
    }
    \caption{Homepage of \pcxg displaying the user's datasets, annotations, and
    compiled analysis notebooks.}
    \label{fig:screenshot}
\end{figure}

\begin{figure}[htbp]
    \centering
    {
        \setlength{\fboxsep}{0pt}
        \setlength{\fboxrule}{1pt}
        \fbox{\includegraphics[width=\textwidth]{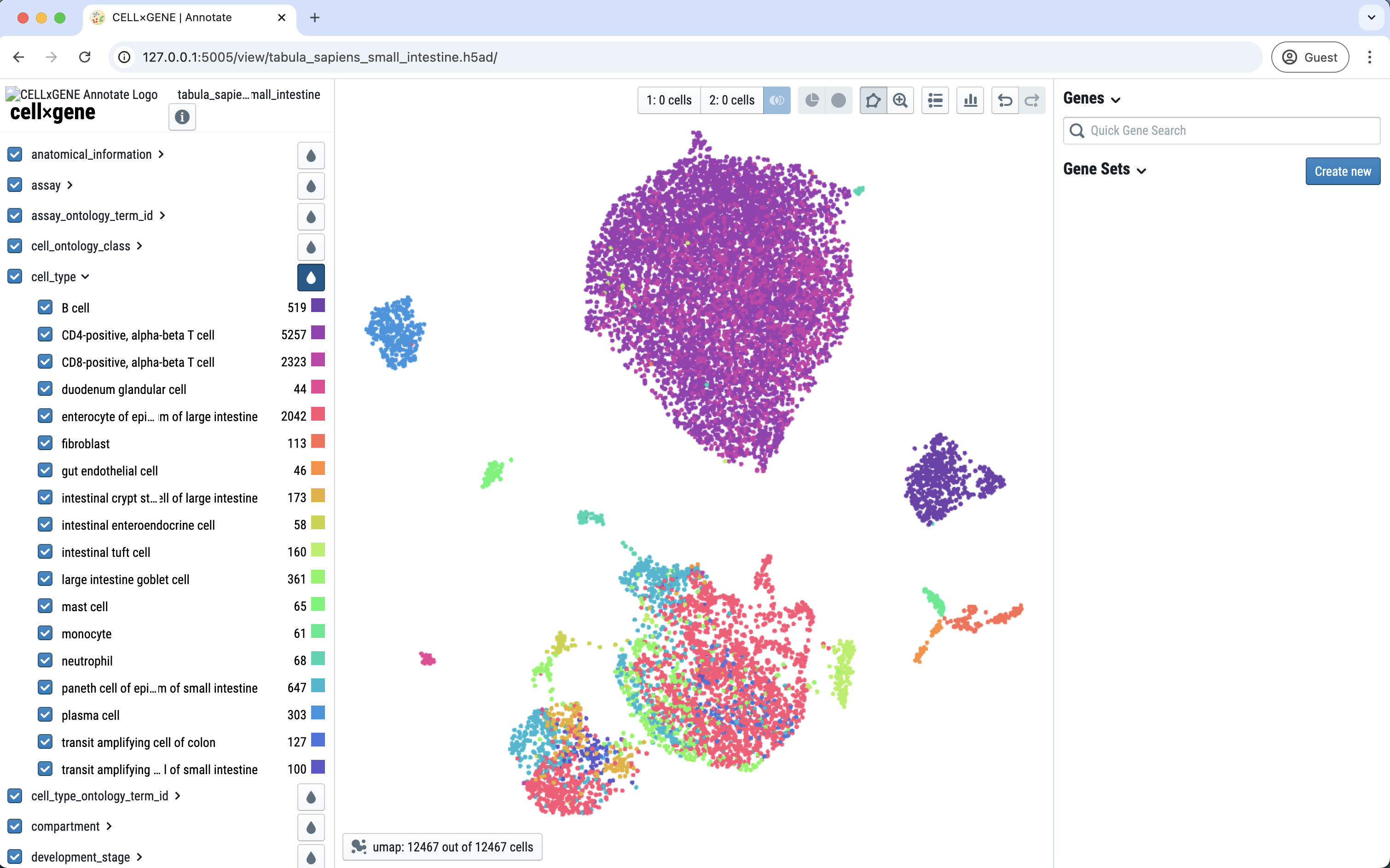}}
    }
    \caption{Screenshot of the CELLxGENE interface available within \pcxg.}
    \label{fig:cellxgene_interface}
\end{figure}

The homepage of the app is a slightly modified version of the CELLxGENE-Gateway
file browser, designed to improve usability (Figure~\ref{fig:screenshot}).
From this page, datasets can be loaded for analysis in CELLxGENE. Sets of
annotations and gene sets can be created or loaded, along with compiled
Python/R notebooks showing how the data has been processed previously. Running
CELLxGENE sessions can be identified and terminated if desired.  Once a dataset
has been loaded through this page, a CELLxGENE instance is launched and can be
used in the standard way, with gene sets and cell annotations being created and
saved as \texttt{csv} files (Figure~\ref{fig:cellxgene_interface}). No internet
connection is required since all software is hosted locally. 

\subsection{Installation}

\begin{figure}[htbp]
    \centering
    \includegraphics[width=0.7\textwidth]{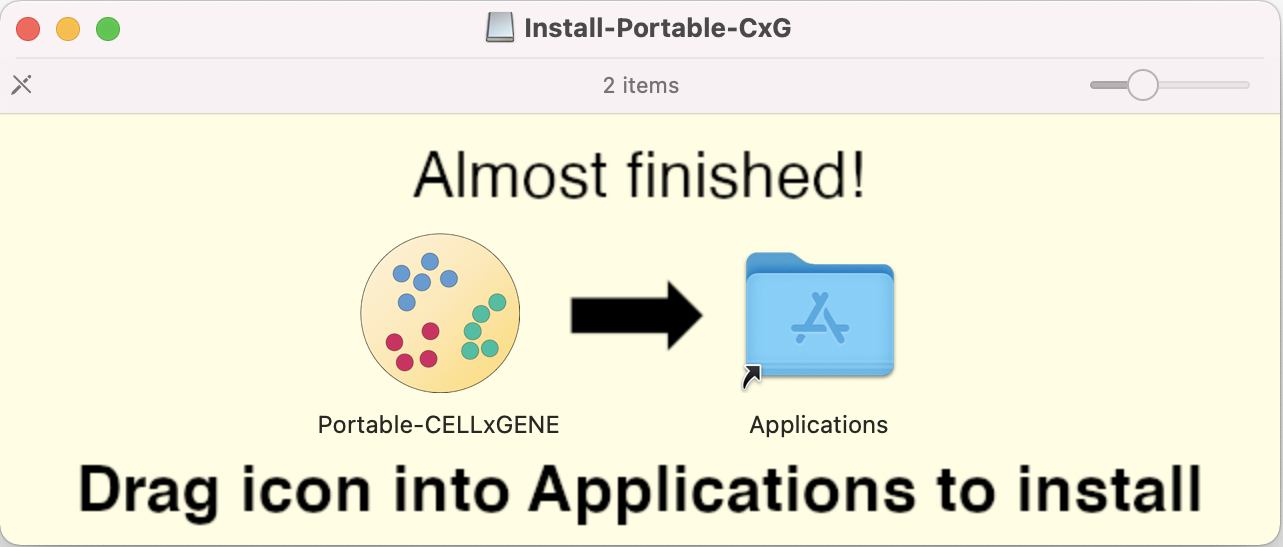}
    \caption{Screenshot of the MacOS \texttt{dmg} installer. The user simply
    drags the app icon into the Applications folder.}
    \label{fig:dmg_installer}
\end{figure}

\begin{figure}[htbp]
    \centering
    \includegraphics[width=0.7\textwidth]{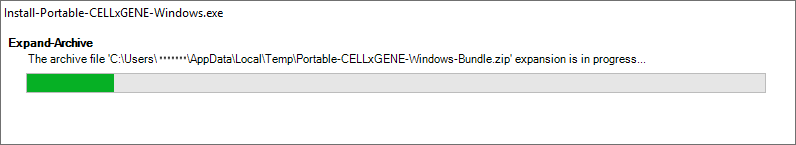}
    \caption{Screenshot of the Windows installer. The user downloads and runs a
    small executable, which installs the software.}
    \label{fig:windows_installer}
\end{figure}

\pcxg can be installed with a straightforward graphical interface on MacOS and
Windows. On MacOS, a standard \texttt{dmg} installer allows the user to simply
drag the application into the \texttt{Applications} directory
(Figure~\ref{fig:dmg_installer}). On Windows, an executable can be run which
downloads and installs the software automatically
(Figure~\ref{fig:windows_installer}).

\subsection{Uninstallation}

\pcxg is similarly easy to uninstall. On MacOS, the application can simply be
deleted from from the Applications directory. On Windows, it can be deleted
from the \texttt{Local Application Data} where it has been installed, along
with its desktop and start menu shortcuts. A script for this process is
available in the GitHub README.

\subsection{Build process}

The build process of \pcxg is designed to be straightforward for both MacOS and
Windows.  It is detailed in the GitHub README.

\subsubsection{MacOS}

Bash scripts to automate the build process are available in the \pcxg GitHub
repository. Briefly, the process comprises two stages: building and signing. In
the building stage, Platypus~\cite{platypus} compiles a MacOS application
containing the conda environment and a script to launch the program. In the
signing stage, the MacOS code signing tool \texttt{codesign} is used to add an
accredited developer identity to the app and all libraries and executables
within the conda environment. An entitlement is added to allow the just-in-time
compilation required by some elements of the \pcxg conda environment (this
ability is by default disabled within the `hardened runtime environment'
necessary for notarization on MacOS). Finally,
\texttt{node-appdmg}~\cite{node_appdmg} incorporates the signed software into a
\texttt{dmg} file for easy installation, which itself is then signed and
released.

\subsubsection{Windows}

Build scripts, assets, and the Windows conda environment are all hosted on the
\texttt{Portable\-CELLxGENE-assets} GitHub repository~\cite{pcxg_assets}.  The
batch script to activate the conda environment and run CELLxGENE-Gateway is
converted into an executable with the \texttt{bat2exe} tool \cite{bat2exe}.
This executable can then be signed with an extended validation certificate.
Finally, an installer is created by converting the powershell installation
script downloaded from the assets repository into an executable using
\texttt{ps2exe}~\cite{ps2exe}. This installer executable can then be signed and
released.

\clearpage

\section{Availability of supporting source code and requirements}

\begin{itemize}
    \item{Project name: Portable-CELLxGENE}
    \item{Project home page: \ghurl}
    \item{Operating system(s): MacOS and Windows}
    \item{Programming language: Python, Bash, Batch, Powershell}
    \item{Other requirements: None}
    \item{License: GNU General Public License v3}
    \item{RRID: \texttt{SCR\_026140}}
\end{itemize}

\section{Contributing guidelines}

Developers are welcome to contribute to Portable-CELLxGENE by submitting a pull
request on the GitHub repository~\cite{pcxg_pull_reqests}. All contributors are
expected to adhere to the Code of Conduct~\cite{pcxg_code_of_conduct}.

\section{Data availability}

Any dataset in the \texttt{h5ad} format can be used in Portable-CELLxGENE.
CELLxGENE Discover~\cite{cxg_discover} is one example of a database of files in
this format. The \texttt{h5ad} file must be stored in a directory which is then
accessed with the folder selection window. Detailed instructions are available
in the README in the GitHub repository.

Three datasets from the Tabula Sapiens~\cite{tabula_sapiens} were downloaded
from CELLxGENE Discover and used to generate the figures in this paper:
\texttt{Large\_Intestine}~\cite{tabula_sapiens_large_intestine},
\texttt{Small\_Intestine}~\cite{tabula_sapiens_small_intestine}, and
\texttt{Pancreas}~\cite{tabula_sapiens_pancreas}.

\section{Abbreviations}

\begin{itemize}
    \item{\emph{GUI}: graphical user interface}
\end{itemize}

\section{Ethical approval}

Not applicable.

\section{Conflict of interest}

None declared.

\section{Funding}

The author is funded by the NIHR Great Ormond Street Biomedical Research Centre.

\section{Acknowledgements}

Thank you to Joe Davidson, Yara Sanchez Corrales and others for testing and
providing feedback on the installation process and design of \pcxg. Thank you
to Sergi Castellano for feedback on the manuscript.

\bibliographystyle{ieeetr}
\bibliography{bibliography}

\end{document}